# Strong Coupling between On Chip Notched Ring Resonator and Nanoparticle

S. Wang[1], K. Broderick[1, 3], H. Smith[1], and Y. Yi[2, 3,1 *]

*[1]Massauchusetts Institute of Technology, Cambridge, MA 02139*
*[2]New York University, New York, NY 10012*
*[3]CUNY Graduate Center, New York, NY 10016*

We have demonstrated a new photonic structure to achieve strong optical coupling between nanoparticle and photonic molecule by utilizing a notched micro ring resonators. By creating a notch in the ring resonator and putting a nanoparticle inside the notch, large spectral shifts and splittings at ***nm*** scale can be achieved, compared to only ***pm*** scale observed by fiber tip evanescently coupled to the surface of microsphere, thereby significantly lowered the quality factor requirement for single nanoparticle detection. The ability for sorting the type of nanoparticles due to very different mode shift and splitting behavior of dielectric and metallic nanoparticles is also emphasized.

**e-mail: yys@alum.mit.edu*

With the rapid progress of nanotechnology, many nano scale photonic devices as small as 30nm have been realized, which are very promising to achieve manipulation of photons at chip scale and having broad applications in renewable energy (photovoltaic cells, solid state lighting), telecommunications and bio medical field[1]. Recently, it was found that the electromagnetic modes of certain photonic devices are very similar to the electronic wave function of molecules[2-3]. One of the most interesting examples is micro ring resonators, especially when we arrange two or more microresonators together within optical coupling regime, the electromagnetic modes of the whole structure are very similar to the bonding (symmetric) or anti bonding (antisymmetric) electronic wave function modes formed in molecules[4-7]. It is interesting to study the photonic molecule of various structures using optical techniques and it may further improve our understanding of the real molecular structures. Nanoparticle has played a key role in nanophotonics and has found many applications in medicine, drug delivery, solar cells and sensors. It is also an important tool for the study of many nano scale structures and is used to interact with nano scale devices, as critical information could be obtained to understand their characteristics. Recently, nanoparticles have been heavily used in the optical sensor area, as fast, non-invasive, and potentially label-free techniques are becoming more important for bio-sensing, gas sensing, chemical sensing. Single nanoparticle detection is one of the ultimate goals for a sensing device and represents sensing at the extreme. In recent years, many novel methods have been utilized to realize nanoparticle detection. For example, metal nanoparticles are used as contrast agents in bio molecule sensing, semiconductor nanoparticles are used as single photon emitters in quantum information processing, and as fluorescent markers for biological processes, nanoshells with special engineering methods are used for cancer therapies and photothermal tumor ablation, polymer nanoparticles are employed as calibration standards and probes in biological imaging in functionalized form[8-11]. The synergy between the photonic molecule and nanoparticle provides us with a unique opportunity, as we can utilize the special photonic molecular modes to interact with the nanoparticle to achieve single nanoparticle detection, inversely, we can also utilize nanoparticles to study photonic molecules and their properties.

In this work, we used on chip notched micro resonators as a photonic molecular example, as a variety of types of optical micro-resonators have been investigated and is a natural photonic molecule to use (Fig.1a); for the nanoparticle, we used AFM tip to simulate a single nanoparticle, where the small tip can be either dielectric materials (Si, GaAs, $Si_3N_4$, etc) or metallic materials (Au, Ag, Al, etc). We have achieved, for the first time, strong coupling between an on-chip notched microring resonator and a single nanoparticle. Specifically, we have used the nano scale notch (~100nm) in the micro ring resonator with diameter around 4μm to strongly interact with the Atomic Force Microscope (AFM) tip. The intentionally created notch in the ring resonator cause the splitting of the original ring resonance mode and formation of bonding (symmetric) photonic states and anti bonding (anti symmetric) states. The AFM tip can be positioned inside the notch. The strong coupling between core electromagnetic modes in the notch and the tip cause the bonding photonic modes to shift in ***nm*** scale, while there is almost no shift for the anti bonding photonic modes. It confirms the photonic molecular modes characteristics generated by the notched micro resonator. The result suggests the potential to deeply study the photonic molecular mode characteristics. Furthermore, we have found the unique and very different shift behavior of the splitting modes from the dielectric Si tip and the metallic Au tip, which can be a critical detection and sorting mechanism for the different type of nanoparticle systems.

In the conventional approach to study the splitting modes using microspheres, because the interaction between the fiber tip (or nanoparticle) and the evanescent mode of the microsphere is very small, it leads to a very small change in the effective refractive index of the sphere, which shifts the wavelength position of the peaks and causes small splitting in the resonator transmission (or reflection) spectrum, these changes are typically on the order of ***pm*** in size[12]. In order to detect such small shifts, one must normally use an expensive tunable narrow-linewidth laser source to scan the relevant spectral region of the resonator output spectrum. Furthermore, the resonator itself must be designed to yield a very narrow linewidth, so that the small peak shifts and splitting can be detected. This requires a high finesse (free spectral range divided by linewidth), or equivalently high quality factor (operating wavelength divided by linewidth ~ $10^8$), which translates to low loss waveguides in the resonator and weak coupling between the resonator and fiber tip. Here we demonstrated strong coupling between on chip notched ring resonator and nanoparticle, where the nanoparticle can be placed inside the notch. Compared to the ring resonator without notch, as illustrated in Fig.1b, where the interaction strength between the nanoparticle and the resonator mode field is relatively weak and only a small portion of the field (evanescent tail) is interacting with the nanoparticle, the notch provides access to the peak of the electromagnetic field localized in the core, so that when a nanoparticle is placed there, the strong core field, rather than the weak evanescent cladding field, overlaps the nanoparticle and thereby produces an enhanced response, as shown in Fig.1c.

The on-chip photonic device configuration of a micro resonator with a notch, nanoparticle and integrated with two bus waveguide has been fabricated[13-17]. The notch of 100nm size in the ring was fabricated by e-beam lithography. We have analyzed the case of a 100nm long notch with a 20nm diameter dielectric nano particle Si tip and 20nm diameter metallic Au particle inside the notch. The thickness of Si waveguide is 220nm. Fig. 2a is the image of the 100nm sized notch in the micro resonator using portable SEM (Hitachi TM-1000). The ring resonator is 4.0μm in diameter with the waveguide width of 200nm. The core material of the ring is Si with refractive index 3.48 at around 1.53μm wavelength, and the bottom cladding material is $SiO_2$ with refractive index 1.46. We used SOI wafer for the small ring resonator fabrication, and an e-beam is utilized to fabricate the 100nmx100nm notch size at the edge of the Si micro ring resonator. The refractive index of the Au nanoparticle tip is $0.54 + 9.58i$ at 1.53μm. Fig.2b is the SEM image within the coupling gap area, which shows the clear 100nm gap between the bus waveguide and the notched ring resonator. A tunable laser from 1480nm to 1580nm is used to couple the light from tapered optical fiber to the Si waveguide, a Ge detector is put at other end to collect the through port signal. Fig.3 is the measurement result for the "Through" port of bus waveguide for the ring resonator with notch at the edge. We can clearly see the splitting of the original resonance modes at around 1.53μm, which represents the bonding photonic mode at shorter wavelength (1.534 μm) and the anti bonding photonic mode at longer wavelength (1.537 μm). The splitting width is almost 3nm; this ***nm*** scale splitting bandwidth is much larger than the ***pm*** splitting bandwidth we normally see for the fiber tip close to the microsphere resonator.

We used a Si nanoparticle tip with 20nm in diameter and put it inside the center of 100nm size notch of the ring resonator using portable AFM, transmission at "Through" port of the coupled waveguide is shown at Fig.4a. It is observed that the bonding mode is ***red*** shifted by 1.1nm, while the position of the anti bonding mode is almost unchanged, leading to a smaller splitting bandwidth due to the red shift of the bonding mode. Next the Si nanoparticle tip is replaced with a 20nm Au nanoparticle tip and also put inside the notch center of the micro ring resonator. We see the dramatically different shift with the Au nanoparticle for the bonding state, a 1nm ***blue*** shift is observed instead and the anti bonding state remains unchanged, causing larger splitting bandwidth than the original splitting due to the notch with just air, as shown in Fig.4b. The results reveal the drastically different photonic mode properties between the symmetric and anti symmetric states and suggests that we can utilize the nanoparticle to study the photonic molecule characteristics. We can understand the red shift induced by the Si nanoparticle by considering an effective index increase in the notch due to the substitution of air with high index Si; the same reason for Au nanoparticle, as the real part is less than that of Air. We can also understand the fact that nanoparticles have little effect on the anti-bonding mode by considering that they are placed at the center of the notch and close to the zero field node of the anti-bonding mode. It can be clearly seen that the strong coupling effect of a nanoparticle placed in a notch is fundamentally different from the evanescent coupling effect of a nanoparticle placed in the evanescent field of the resonator, which high Q microspheres are often used.

The different wavelength shift between a dielectric Si nanoparticle and a metallic Au nanoparticle, as well as the splitting bandwidth narrowing with the dielectric Si compared to the widening with the metallic Au nanoparticle, provides us with a unique, *self referencing* mechanism to distinguish these different types of nanoparticles. This is very important for the bio sensor area utilizing nanoparticles, as Au nanoparticles are often used for tagging, but other dielectric nanoparticles other than the analyte may be present and cause misleading sensing signals, Au nanoparticle's unique effects on the shift direction and the splitting bandwidth widening will enable us to distinguish the real signal from the background dielectric nanoparticle noise and greatly enhance the signal to noise ratio. The intentionally fabricated notch in the micro ring resonator provides us with a localized position to trap the nanoparticle. Based on the strong splitting and wavelength shift when the nanoparticle is localized in the notch, we can identify whether the nanoparticle stays within the fabricated notch and differentiate the nanoparticle position. The future study on the electromagnetic force applied upon the nanoparticle close to the notch will be interesting to help us to understand the mechanic response from the notch and how to deliver the nanoparticle to the notch.

In summary, we have experimentally demonstrated the strong coupling between on chip nanoscale notched ring resonators and nanoparticle, in which a notch is introduced in the resonator to provide access to the core field, which is drastically different from previous studies using evanescent coupling to microsphere. Placing a nanoparticle in the notch produces a much stronger response than simply placing the nanoparticle in contact with the exterior of the core. In the exemplary case of a dielectric silicon and metallic gold nanoparticle placed in a notch, we have demonstrated that nanoparticle induces a large wavelength splitting (~**nm**) and very different shift in the resonant modes of the resonator. This is a significant improvement over the smaller wavelength shifts and splitting (~**pm**) observed in earlier experiments where the particle was placed outside the core of a conventional microsphere resonator, and lowers the requirement for very high Q resonator devices. Note that the utility of this approach is not limited to ring resonators used as examples here, but can be extended to other types of resonator geometries, such as racetracks, and polygons. This work provides us a unique way to achieve single nanoparticle detection and sorting with thousands of times signal enhancement. The nature of the on chip microresonators will also make large scale integration on a sing chip possible.

We thank the support from Microsystems Technology Laboratory and Center for Materials Science and Engineering at MIT for the access to the facility and measurement equipments, High Performance Computing center at CUNY, and 3M central research lab.

**Figure captions:**

**Figure 1** (a) The notched micro ring resonator and the coupling to the bus waveguide with input and output ports. The bus and ring waveguide width is 200nm. The notch size is around 100nmx100nm. (b) The evanescent coupling between the resonance mode and the nanoparticle. (c) The notch in the microring resonator. The nanoparticle is in the notch and coupling to the core of the resonance mode, the black curve represents the single mode behavior for the ring waveguide. The ring waveguide is a sing mode waveguide at 1.53µm.

**Figure 2** (a) The image of the 100nm size notch by portable SEM. (b) The SEM image with clear 100nm coupling gap between the notched ring resonator and the single mode bus waveguide.

**Figure 3** The transmission at Through port around 1.53µm, the splitting is around 3nm, the bonding photonic mode is at shorter wavelength 1.534 µm and anti bonding photonic mode is at longer wavelength 1.537 µm.

**Figure 4** (a) Si nanoparticle tip with 20nm in diameter is put inside the 100nm notch size of the ring resonator, the Through port from the coupled waveguide is shown, it is observed that the bonding mode wavelength (resonance at shorter wavelength) is ***red*** shifted with 1.1nm, while the wavelength of the anti bonding mode (resonance at longer wavelength) is almost unchanged, the splitting bandwidth is smaller due to the red shift of the bonding mode. (b) 20nm Au nanoparticle tip is put inside the notch size of the micro ring resonator, we see the dramatic different shift from the Au nanoparticle for the bonding state (resonance at shorter wavelength), the 1.0nm ***blue*** shift is observed instead and the anti bonding state (resonance at longer wavelength) is still unchanged, the splitting bandwidth is larger than the original splitting due to the notch with air.

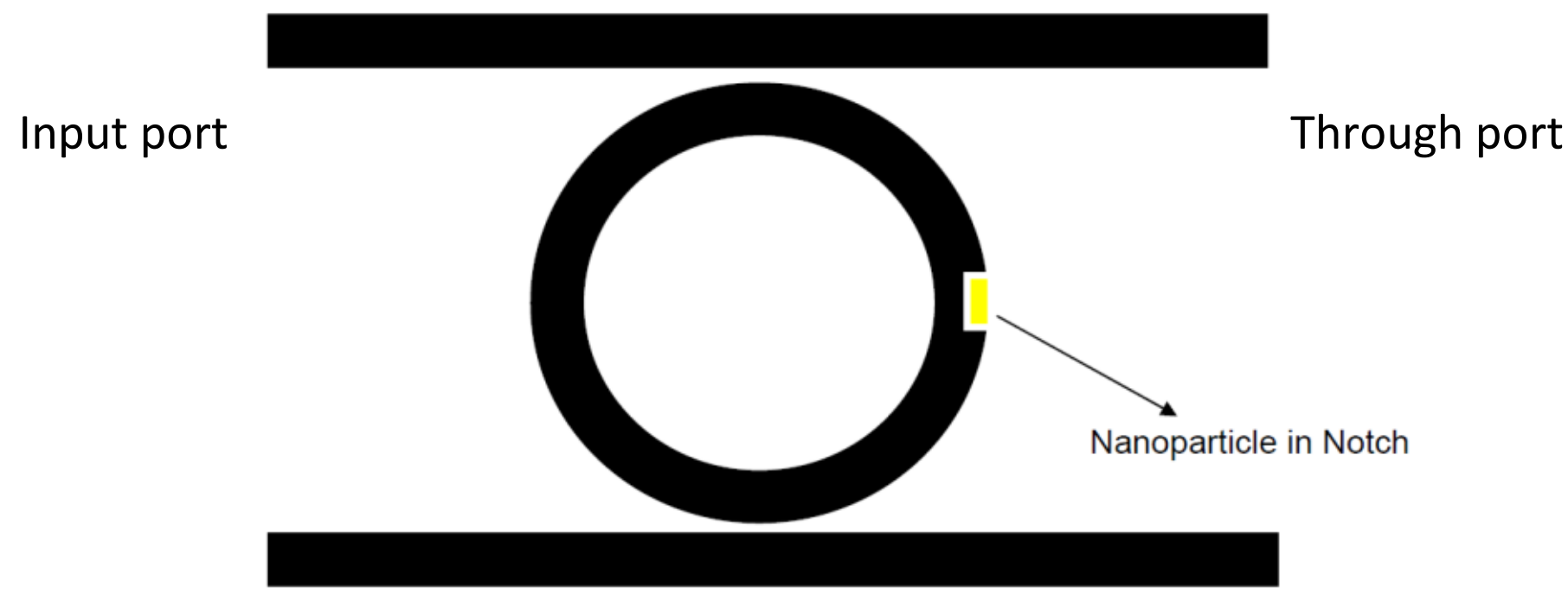

(a)

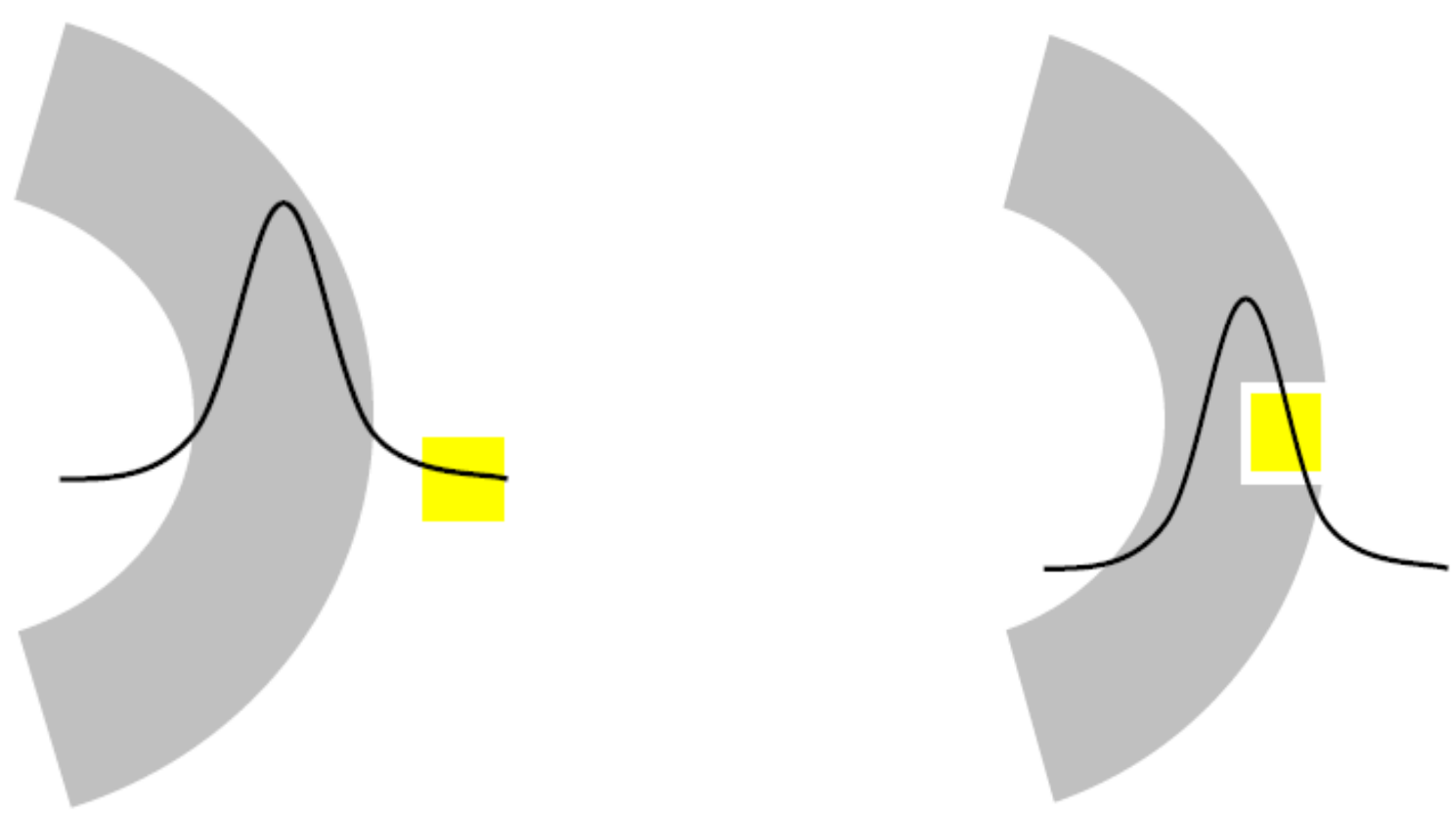

(b) (c)

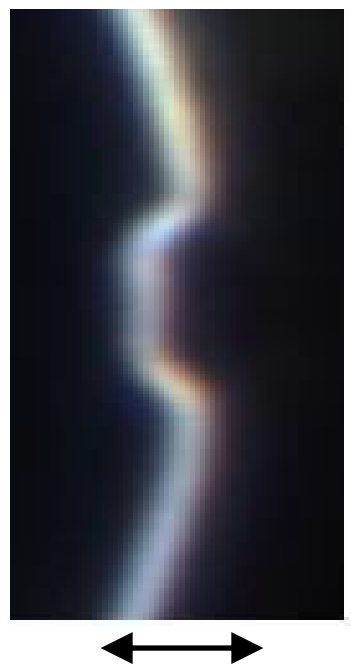

100nm

(a)

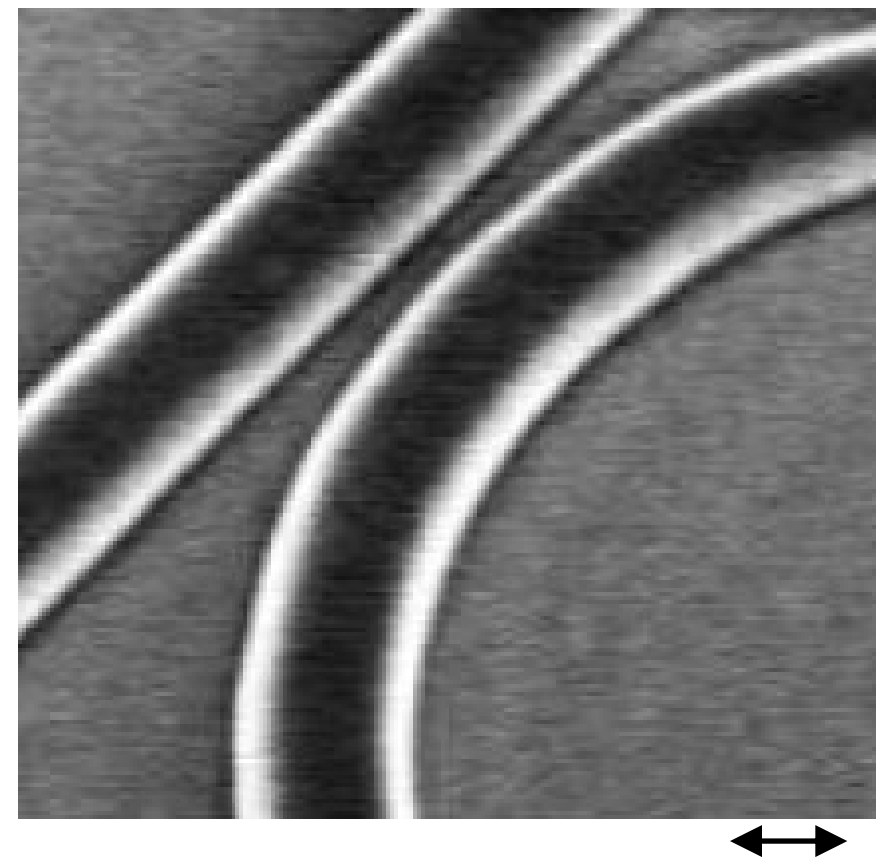

200nm

(b)

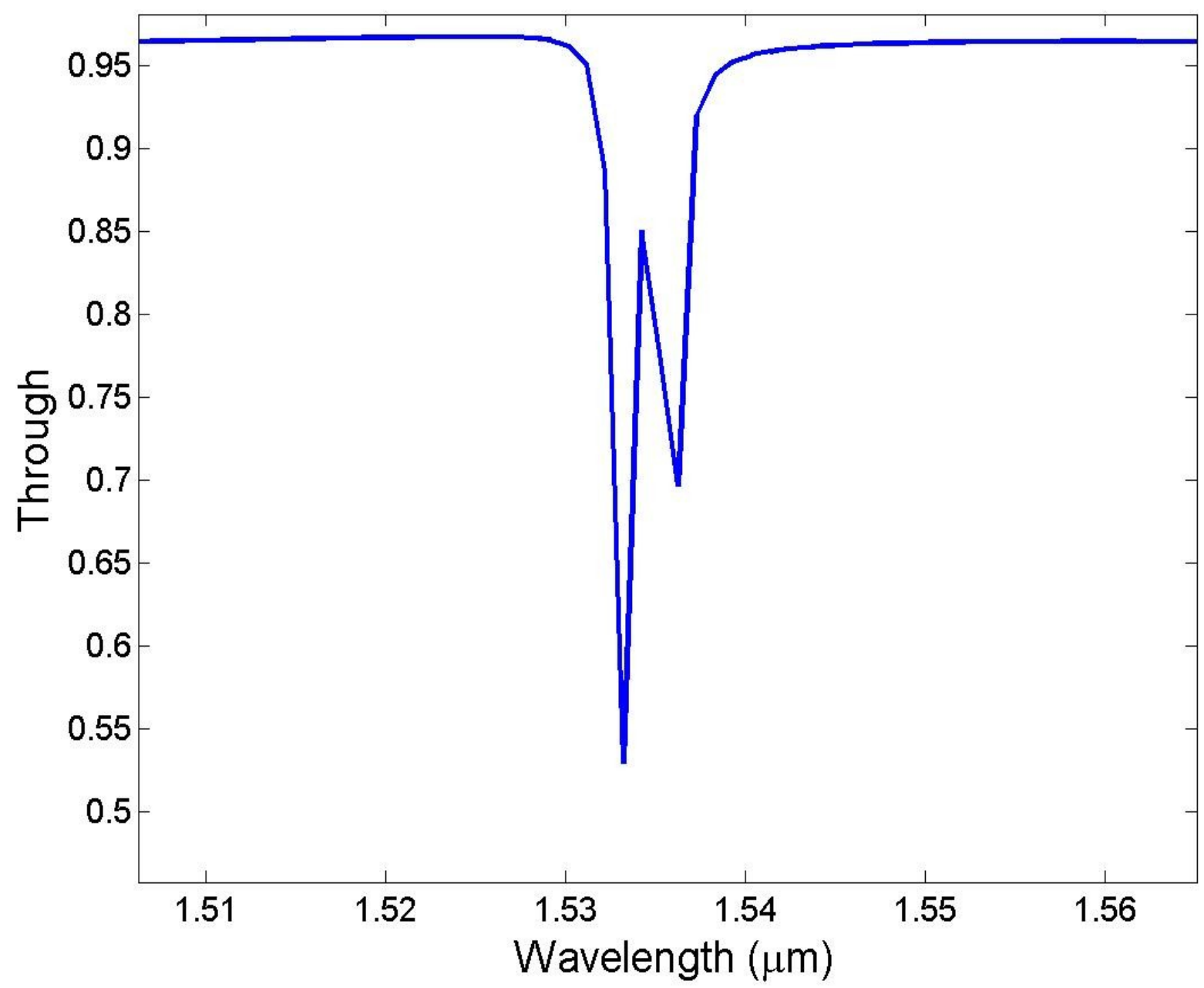
0.95
0.9
0.85
0.8
0.75
0.7
0.65
0.6
0.55
0.5
Through
1.51
1.52
1.53
1.54
1.55
1.56
Wavelength (μm)

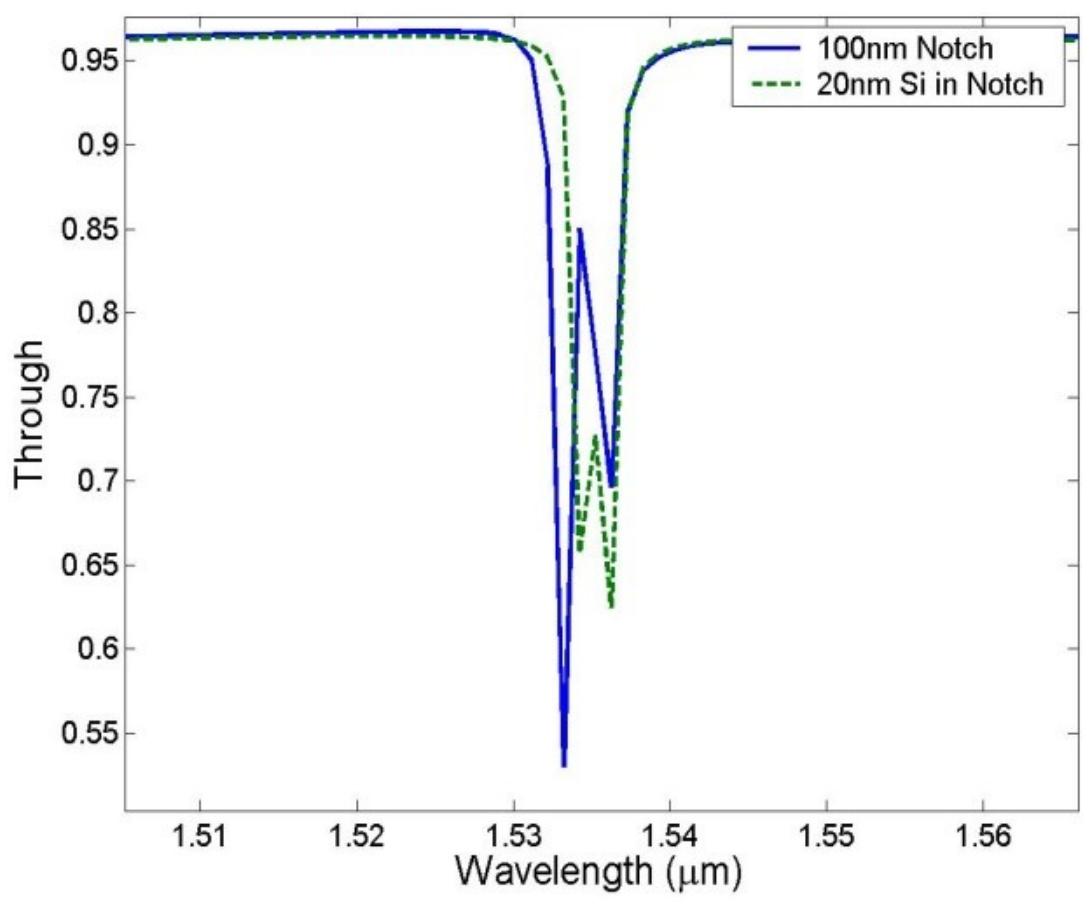
100nm Notch
20nm Si in Notch
Through
Wavelength (μm)
0.95
0.9
0.85
0.8
0.75
0.7
0.65
0.6
0.55
1.51
1.52
1.53
1.54
1.55
1.56

(a)

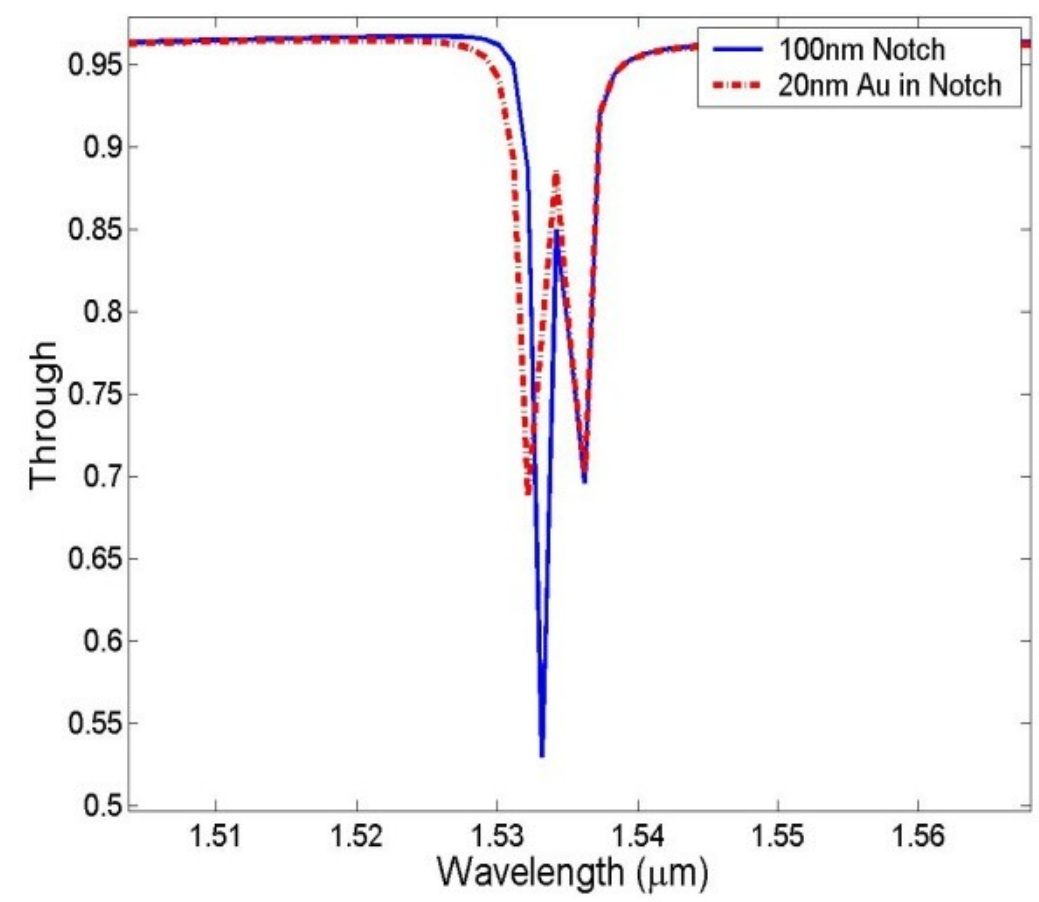
100nm Notch
20nm Au in Notch
Through
Wavelength (μm)
0.95
0.9
0.85
0.8
0.75
0.7
0.65
0.6
0.55
0.5
1.51
1.52
1.53
1.54
1.55
1.56

(b)